# A Near Infrared Laser Frequency Comb for High Precision Doppler Planet Surveys

S. Osterman[1,a], S. Diddams[2], F. Quinlan[2], J. Bally[1] J. Ge[3] and G. Ycas[2,4]

[1]Center for Astrophysics and Space Astronomy, University of Colorado, Boulder, CO 80303, USA
[2]Time and Frequency Division, National Institute of Standards and Technology, Boulder, CO, 80305, USA
[3]Bryant Space Science Center, University of Florida, Gainesville, FL 32611, USA
[4]Department of Physics, University of Colorado, Boulder, CO 80309, USA

**Abstract.** Perhaps the most exciting area of astronomical research today is the study of exoplanets and exoplanetary systems, engaging the imagination not just of the astronomical community, but of the general population. Astronomical instrumentation has matured to the level where it is possible to detect terrestrial planets orbiting distant stars via radial velocity (RV) measurements, with the most stable visible light spectrographs reporting RV results the order of 1m/s. This, however, is an order of magnitude away from the precision needed to detect an Earth analog orbiting a star such as our sun, the Holy Grail of these efforts. By performing these observations in near infrared (NIR) there is the potential to simplify the search for distant terrestrial planets by studying cooler, less massive, much more numerous class M stars, with a tighter habitable zone and correspondingly larger RV signal. This NIR advantage is undone by the lack of a suitable high precision, high stability wavelength standard, limiting NIR RV measurements to tens or hundreds of m/s [1,2].

With the improved spectroscopic precision provided by a laser frequency comb based wavelength reference producing a set of bright, densely and uniformly spaced lines, it will be possible to achieve up to two orders of magnitude improvement in RV precision, limited only by the precision and sensitivity of existing spectrographs, enabling the observation of Earth analogs through RV measurements. We discuss the laser frequency comb as an astronomical wavelength reference, and describe progress towards a near infrared laser frequency comb at the National Institute of Standards and Technology and at the University of Colorado where we are operating a laser frequency comb suitable for use with a high resolution H band astronomical spectrograph.

## 1 Introduction

To date, over 300 planets have been discovered around nearby stars, with the number constantly increasing [3,4]. RV studies measuring the Doppler shifts in the host star induced by the orbiting planet are responsible for identifying the overwhelming majority of the exoplanets now known. These detections include several multi-planet systems, planets with masses as low as 5 $M_{Earth}$, and planets with orbits extending beyond 5 AU [1,5]. Numerous planned ground- and space-based projects will be coming on line in the next two decades to expand this effort. Planet finding strategies will include transit detections, interferometric astrometry, and direct imaging. Even with a welcome diversity of discovery methods, RV studies will remain a key component of exoplanet detection and characterization. In addition to finding new planets, RV studies allow detection of multiple planets in a single system and facilitate determination of exoplanetary system parameters such as mass, metallicity, and orbital eccentricity.

---

[a] e-mail : isaline.boulven@edpsciences.org

The detection of Earth-mass planets in stellar habitable zones is one of the principal goals of exoplanet studies. Cool stars represent promising systems in which to detect habitable Earth-like planets: Low mass M-type stars are much more numerous than their more massive counterparts and their reduced temperature moves their habitable zones closer in, resulting in a larger RV signature for a terrestrial planet in an M star's habitable zone compared to more massive stars. For example, while an Earth/Sun analog system would exhibit ~10 cm/sec RV shift over a one year orbit, an earth mass planet orbiting an M6V dwarf at the inner edge of its habitable would induce a 2 m/s signal over its 3 day orbit, and a 5 $M_{Earth}$ planet would induce a 10m/s RV signature (Fig. 1) [6,7,8,9]. See, for example, the two potentially habitable super earth mass planets orbiting M3 dwarf Gliese 581 [10,11]. Although M3 and earlier objects can be studied at optical wavelengths, the near-infrared (NIR) has the advantage that stellar fluxes peak above 1 μm, increasing the number of observable systems, while spectral-line features remain sufficiently abundant for RV studies [12,13]. M dwarfs can be 10-100 times brighter in the NIR than in the visible, sot that a Doppler search in the near-IR using spectrographs with resolution and precision comparable to those currently used in the optical could observe targets as faint as J~10-12 mag (V=16-18 mag). At these magnitudes there are perhaps a thousand times more M dwarfs available for observations than at V~11 mag. [14,15]. Despite all this, these advantages cannot be fully exploited due to the lack of a suitable wavelength standard [2].

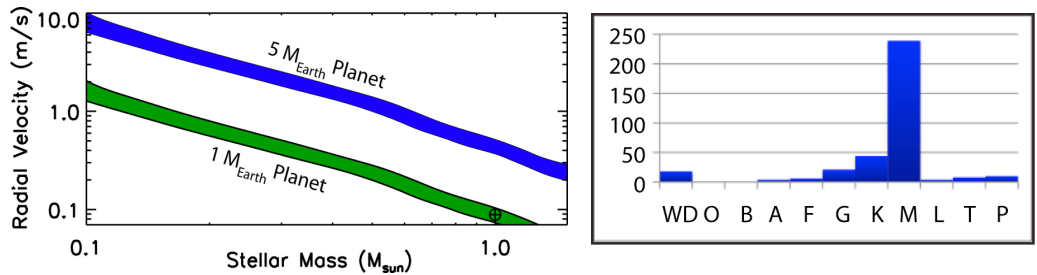

**Fig. 1.** Left: Stellar RV amplitude for a terrestrial planet (1$M_{Earth}$ and 5$M_{Earth}$) orbiting within the habitable zone as a function of stellar mass. Earth is indicated in the lower right (after Kasting [8]). Right: Distribution of stars by spectral class within 10pc showing the predominance of cooler M stars [16].

## 2 Calibration standards

As planet finding methods become more mature and sophisticated, many of the current limitations - including longer time baselines and improved stellar noise subtraction, as well instrumental effects such as fiber induce noise, thermal stability, and sensitivity - are being addressed through continued telescope and spectrograph refinement, pointing to the development of a high precision velocity calibration source as a critical requirement for improved RV precision, especially in the NIR.

### 2.1 Current calibration standards

Wavelength calibrations sources can be grouped into two categories: Emission and absorption. For emission sources, an ideal wavelength standard would provide a dense array of uniformly bright, uniformly spaced narrow lines. Line density could be tailored to the specific spectrograph in use (ideally ~ 1 line per 3 resolution elements) and each line could be traced to fundamental constants.

To date, the primary wavelength standards for the UV to NIR wavebands have been line sources such as PtNe or ThAr lamps and absorption cells, most notably iodine cells. While these have proven to be of enormous value, they have significant limitations. In many respects, the absorption cell provides an outstanding wavelength standard, with the calibration spectra exactly superimposed on the science optical path; however, absorption cells suffer from several factors that limit its utility in the NIR. The iodine cell [18] is restricted to a limited wavelength range (500-630nm), limiting the number of available targets. In addition, the absorption cells superimpose their absorption profile on the stellar spectrum and attenuate continuum flux, resulting in significant loss in observed target counts. Moreover, because the calibration source is in absorption, the observed spectrum must be of high signal-to-noise (≥200) to extract the highest precision measurements. As a result, iodine cell based RV observations are restricted to bright visible sources, making the iodine cell ill suited for observing dim objects with peak spectral brightness in the NIR. Although M stars dominate the stellar population of the Milky Way, relatively few are under observation per program due to their low brightness at optical wavelengths [3,18].

The use of alternative fill gases has been explored for use in the NIR. Absorption cells filled with hydrogen halides such as HCl, HI, and HBr provide in moderately dense coverage over ~ 50% of the 1.0-2.0μm band pass, but with widely spaced lines whose depths vary by an order of magnitude in each system [19]. For example, the $H^{35}Cl/H^{37}Cl$ bands at 1745nm have lines spaced at 1 to 5 lines per 10nm [20] providing one line for every 60 to 300 resolution units of an R=50,000 spectrograph, compared to the ideal spacing of one line every three resolution elements. In several regions, these cells leave absorption feature free gaps of up to 100 nm, creating vast spectral wastelands devoid of reference lines. Work is also in progress to develop a set of absorption cells using $C_2H_2$, $^{12}CO$, $^{13}CO$ and HCN for use in the H band, providing somewhat better line density (up to 20 lines per 10nm), but only from 1.51 to 1.63 μm [21].

The simultaneous reference technique injects an emission wavelength calibration source into the spectrograph offset from the science light path in the spatial direction so that science and calibration spectra are interleaved in the image plane and may be recorded simultaneously. This approach has already achieved observing precisions on the order of 1 m/s and led to the discovery of a host of planets [5,22]. Murphy et al. carried out a detailed study to improve the characterization of a ThAr lamp and were able to achieve an average 1 m/s precision [6,23]. Despite this, the ThAr simultaneous reference technique is coming up against fundamental performance limitations including the presence of unresolved line blends, uncertainties in line identifications and locations, widely varying and generally low intensities and low line densities at key wavelength regions (most notably the NIR for ThAr lamps). In addition, each lamp is unique, with lamp to lamp variability due to cathode and fill gas impurities, age and window condition.

Alternative emission line sources (hollow cathode and pen ray lamps using a variety of metals and Nobel gases) have been found to be viable calibration sources, but the useable spectrum is as sparse as absorption cell spectra, with widely varying and generally low overall intensity and substantial regions lacking in usable lines. ThAr lamps have a cluster of bright lines at 1.3 μm, but are much less useful at longer wavelengths [22,24]. If the observer is fortunate enough to have the available lines correspond to the spectral range of interest, then these lamps can provide a high degree of precision, having supported solar observations with RV precision as low as 10 m/s in the very near IR and beyond 2.0 μm, but not at intermediate wavelengths.

Finally, NIR observations may be calibrated through observation of telluric lines. These are typically limited to 10-25m/s precision, are highly variable in depth and velocity structure from night to night and between observing sites and are restricted to wavelengths shorter than 1.1μm and beyond 2.0μm [25].

## 2.2 Laser frequency comb technology

In the NIR, absorption and emission sources are plagued by all of the limitations of their visible light counterparts without the heritage that comes from years of refinement. Specifically, developing a pipeline that can take into account the signal loss introduced by an absorption cell and that will reliably determine the exact centroid of a reference line when superimposed on a science spectrum, or knowing which of the limited array of reference lines are reliably stable, unblended and insensitive to lamp aging or lamp-to-lamp variation will take years of development, and in the end be limited to no better than the state of the art in the visible (~1m/s). In contrast, by producing an array of uniformly bright, evenly spaced lines with absolute wavelength knowledge traceable to the standard second, a laser frequency comb can support <<1m/s precision [26,27].

The use of a mode-locked laser as a tool for optical frequency metrology was first demonstrated with picosecond lasers in the late 1970's, and was re-introduced nearly three decades later when Hänsch and co-workers were able to stabilize an octave spanning comb by self referencing [28,29]. Hänsch et al. showed that the comb of frequencies emitted from a mode-locked laser can be used as a precise "optical frequency ruler" where the n$^{th}$ mode $f_n$ of the laser frequency comb is described by the simple and exact expression

$$f_n = f_0 + nf_r \quad (1)$$

where the carrier envelope offset, $f_0$, and the pulse repetition rate, $f_r$, are on order of a few hundred MHz to a few GHz. A pulse train can be related to the frequency domain as

$$E(t) = \text{Re}\left(A(t)\exp(-i2\pi f_c t)\right) = \text{Re}\left(\sum_n A_n \exp(-i2\pi(f_0 + nf_r)t)\right) \quad (2)$$

with the carrier frequency given as $f_c$. This is illustrated in Fig. 2 below.

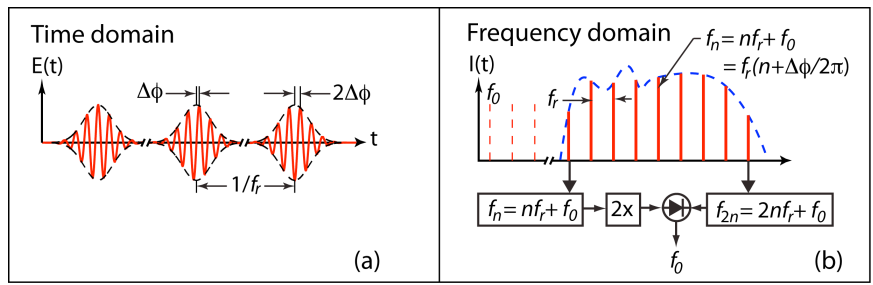

**Fig. 2** In a mode-locked femtosecond laser pulses are emitted at the rate $f_r$. Due to dispersion in the laser cavity, the carrier advances with respect to the envelope by $\Delta\phi$ from one pulse to the next, resulting in an offset common to all modes of $f_0 = f_r\Delta\phi/(2\pi)$. "f-2f" self-referencing allows precise measurement and control of the carrier envelope offset frequency $f_0$ [30].

The carrier envelope offset frequency $f_0$ arises from the difference between the pulse and group phase velocities in the laser. The repletion rate $f_r$ can be readily measured with a fast photodiode, and $f_0$ is measured using the f-2f self-referencing technique shown in Fig. 2b. While Eq. 1 is an exact relationship for $f_0$ and $f_r$, shown to be valid with precision reaching one part in $10^{19}$, lower precision field measurements using the signal from the global positioning system (GPS) as a reference can support accuracies of one part in $10^{11}$, a level of precision that enables cm/s RV measurements [31].

The left panel of Fig. 3 shows frequency comb spectra obtained with three mature, frequency stabilized femtosecond (fs) laser combs spanning the spectral range of ~ 400 nm - 2000 nm. Ti:sapphire lasers span 400-1200 nm, generating combs with mode spacing ($f_r$) ranging from 0.1 to 10 GHz (increased mode spacing corresponds to more widely spaced modes, increasing the suitability for use with moderate to high resolution spectrometers). Yb:fiber and Yb-doped crystal combs span 500-1700 microns with non linear broadening at mode spacings up to 300 MHz, and Er:fiber lasers with non-linear broadening produce comb spectra between 1000 and 2200 nm at mode spacings up to 250 MHz.

One key to harnessing this technology is mode filtering to reduce the mode density of the native comb to a level that can be resolved by current or proposed astronomical spectrographs (Fig. 3, right) [33,34]. For example, an Erbium fiber laser with a repetition rate of 250MHz would produce a line spacing of 0.002 nm at 1.55 μm. By passing this dense spectrum through a carefully designed, precisely stabilized Fabry-Pérot cavity which transmits only 1 line in 50, the line spacing is increased to a level that is suitable for a λ/Δλ=50,000 spectrograph.

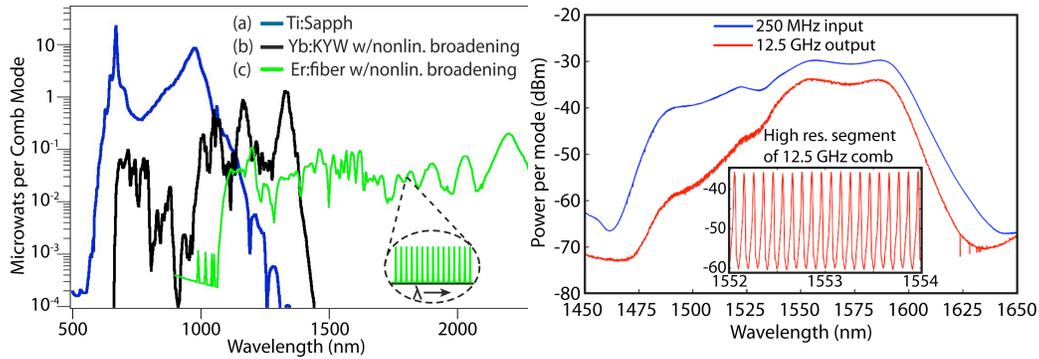

**Fig 3.** Left: Wavelength coverage of different frequency comb technologies. Ti:Sapph lasers can operate at repetition rates ($f_r$) of up to 10GHz. Yb:KYW and Er:fiber lasers operate in the 250MHz range. Right: Unresolved spectrum of the unfiltered 250MHz passively mode-locked Er fiber laser and the 12.5 GHz filtered signal. Increased power loss below 1540nm and above 1610nm due to dispersion in the filter cavity [32].

## 3 The laser frequency comb as a wavelength standard for astronomical spectrographs

While the laser frequency comb appears to be an ideal wavelength standard, there are a number of operational considerations that complicate implementation. Chief among these are filter cavity to source comb detuning and filter bandwidth. On the other hand, the high power output and flexibility of laser combs can overcome obstacles facing other calibration technologies such as source to spectrograph coupling and simultaneous support of multiple instruments.

### 3.1 Cavity detuning, nearest neighbor suppression and bandwidth

The repetition rates of NIR laser frequency combs are too low to produce spectra resolvable by the current generation of NIR astronomical spectrographs. Increased mode spacing is achieved with Fabry Pérot cavities (Fig. 4a), but these introduce their own complications. The mode spacing for a high finesse Fabry Pérot cavity departs from idealized behavior due to wavelength dependent phase shifts introduced in the high reflectivity mirror coatings. This results in increasing misalignment between the frequency comb modes and the cavity transmission peaks towards the band edges.

Because neighboring modes are unresolved by the spectrograph, this detuning will result in an apparent shift in wavelength through two mechanisms: First, because the native comb mode has finite width, the mode will be asymmetrically reshaped as the filter cavity transmission peak moves off of the comb mode. Second, the nearest neighbors to the red and blue side of the central mode will be differentially suppressed, again resulting in an apparent bias in the observed wavelength (Fig. 4b) [34]. This results in a drop in the transmission of the filter cavity which can be used as a diagnostic for magnitude of the misalignment (Fig. 4c).

Even if this detuning is stable with time, it is still undesirable because it impacts one of the chief advantages of the laser comb: Traceability. If the apparent wavelengths of outlying modes departs from the true mode spacing, then this implies that the recorded wavelength is dependent on the mirrors used in the filter cavity rather than being easily traced to fundamental constants. As a result, the spectra recorded with one comb may not be easily reproduced or traced to spectra obtained with different combs. Furthermore, such a departure from the nominal comb relation may bring into question measurements taken with the same comb on the same spectrograph at different epochs or with the same comb on different spectrographs with different line spread functions. While these effects are small when compared to the uncertainties implicit in other wavelength calibration sources, they do represent complications that should not be overlooked.

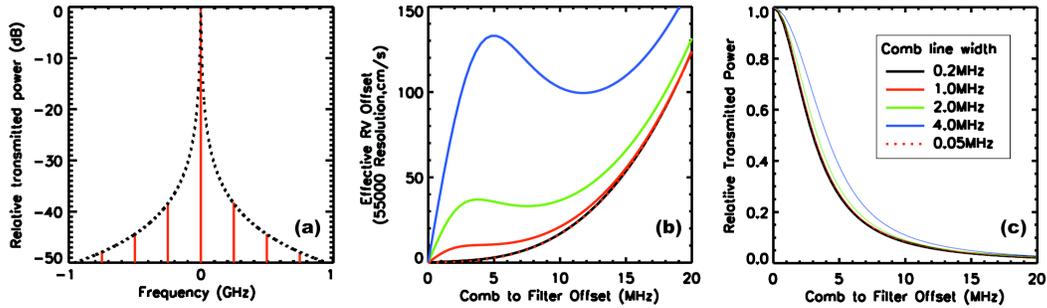

**Fig. 4.** (a) Modeled filter cavity transmission with transmitted comb spectrum overplotted. (b) Predicted RV error induced by cavity to comb detuning as a function of comb line width for a 250MHz comb filtered to 12.5GHz (50:1 filter ratio) and 37dB nearest neighbor suppression for an R=55,000 spectrograph. (c) Corresponding decrease in transmitted power. Note that with either a GPS or rubidium clock comb line widths less than 1MHz should be possible.

### 3.2 Frequency comb signal advantage

There are a number of secondary advantages inherent in the laser frequency comb. The output is readily delivered via optical fibers, and the comb output is exceptionally bright (on order of 100 nW/mode), allowing for the use of very lossy delivery methods. For example, ideally one would deliver calibration light to the spectrograph via a uniform, isotropic source. Two methods to achieve this are dome illumination and slit (or fiber) illumination via an integrating sphere. The best studied line source, the ThAr lamp, is not bright enough in the NIR to support either delivery method [25]. As an example, for a typical, small integrating sphere fed directly from a ThAr lamp and exiting to a 150 μm diameter fiber, the transmission of an integrating sphere will be on order $10^{-6}$ (ignoring fiber coupling losses). With output from the comb on order of 100 nW per mode, there is ample light to accommodate such a high loss coupling mechanism.

In addition to easily coupling into a slit or fiber fed spectrograph, a single comb could be used to support multiple instruments. For example, a NIR 12.5 GHz comb could have its output split, with one feed going to support a 55,000 resolution, H Band spectrograph, and the other feed going to a filter that would pass only 1 in 100 modes from the native comb. While 100:1 filtering with adequate suppression appears on the surface to be impractical, the additional stage in fact only needs to perform 2:1 filtering on the already filtered 12.5GHz comb. Such a signal could then be used to simultaneously support a second, lower resolution instrument. Along similar lines, a small fraction of the native, unresolvable comb signal could be delivered to the spectrographs as a flat field source.

### 3.3 Sample application: FIRST spectrograph at the Apache Point Observatory

As an application example, we consider providing a wavelength calibration signal to the University of Florida's FIRST spectrograph (Fig. 5) [16]. The FIRST instrument is designed around a silicon immersion grating, and is intended to deliver R=55,000 between 1.3 and 1.8μm. FIRST will be fiber fed from the Nasmyth port on the Apache Point 3.5m and a passively mode-locked Er fiber laser will be used to produce a 250MHz comb which is filtered to a mode spacing of 12.5GHz (50:1) for wavelength reference (~0.1nm between transmitted modes at 1550 nm resulting in a relative line spacing of $\lambda/\Delta\lambda$ = 15,500). This signal is fed to the spectrograph via two paths: One fiber (calibration fiber) will illuminate the science fiber feed at the Nasmyth port via a deployable fold mirror. The other (reference fiber) will lead directly to the spectrograph slit plane and inject light into the spectrograph in a position offset from the science fiber in the cross dispersion (spatial) direction. The calibration fiber feed will illuminate the FIRST instrument before and after each exposure and mimic as closely as possible the illumination geometry of the telescope. The reference fiber provides a simultaneous reference spectrum falling in the inter-order spaces of the science spectrum during observations. While feed from the telescope to the spectrograph would likely use a multimode fiber in order to maximize throughput, single mode fibers could be used for the calibration and reference fibers to avoid introducing modal noise in the comb spectrum.

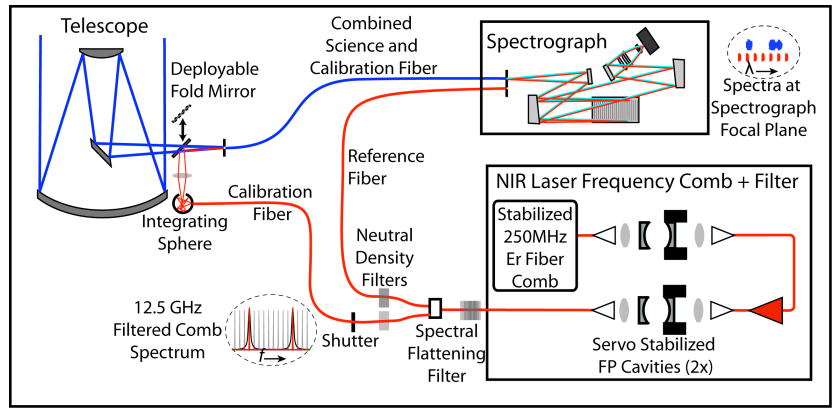

**Fig. 5.** Block diagram of operational setup. The frequency comb provides both simultaneous reference spectra and inter-observation calibration spectra via two separate fiber feeds. Spectra at the focal plane illustrates part of one order with the reference spectrum offset in the spatial direction from the science spectrum.

### 3.4 The CU/NIST laser frequency comb program

An ongoing development program at the University of Colorado and NIST has produced a laboratory frequency comb for use in the NIR with the objective of performing observations with an astronomical spectrograph in the near future. The development program is focusing on three aspects of comb development: Optimal nearest neighbor suppression in the filtered comb spectrum, increased spectral coverage, and spectral flattening. All of these represent critical milestones for an operational wavelength standard.

We have achieved 50:1 filtering of a 250 MHz passively stabilized Er:fiber laser, with >57dB suppression measured between 1525-1580nm (Fig. 3, right panel). [32]. The filtered spectrum shows >10dB variation in output power per mode. Such variability would limit the maximum signal per integration period at the edges of the band, reducing the ultimate signal to noise and centroid precision. By attenuating the brighter portions of the spectrum, a more uniform mode intensity can be achieved. Work is underway at NIST to flatten the spectrum using a spatial light modulator which can provide several decades of modulation in several hundred bins across the comb bandpass, and which can also provide custom modulation to correct for spectrograph throughput variation, ensuring a flat comb spectrum.

Post filter signal amplification and pulse compression, as well as techniques to deliver increased spectral coverage are currently being explored at NIST, with the goal of supporting full H band spectroscopy.

## References

1. S. Udry, X. Bonfills, X. Delfosse, T. Forville, M. Mayor, C. Perrier, F. Bouchy, C. Lovis, F. Pepe, D. Queloz, J.-L. Bertaux, A&A **469**, L43 (2007).
2. L. W. Ramsey, J. Barnes, S. L. Redman, H.R.A. Jones, A. Wolsczczan, S. Bongiorno, L. Engle, J. Jenkins, PASP **120**, 887 (2008).

3.  R. P. Butler, J. T. Wright, G. W. Marcy, D. A. Fischer, S. S. Vogt, C. G. Tinney, J. R. A. Jones, B. D. Carter, J. A. Johnson, C. McCarthy. A. J. Penny, ApJ, **646**, 505 (2006).
4.  J. Schneider, Interactive Extra-solar Planets Catalog, http://exoplanet.eu/catalog.php, accessed 29 October 2008.
5.  S. Udry, G. Torres, B. Nordström, F. C. Fekel, K. C. Freeman, E. V. Glushkova, G. W. Marcy, R. D. Nathieu, D. Pourbaix, C. Turon, T. Zwitter, arXiv:0810.4538v1, to appear in Trans IAU, XXVIIA (2008).
6.  C. Lovis, F. Pepe, F., Bouchy, F., Curto, G.L., Mayor, M., Pasquini, L., Queloz, D., Rupprecht, G., phane Udry, S., and Zucker, S., Proc. SPIE **6269**, 62690P1 (2006).
7.  M. Joshi, AsBio, **3**, 415 (2003).
8.  J. F. Kasting, D. P. Whitmier, R. T. Reynolds, Icarus **101**, 108 (1993).
9.  J. F. Kasting, D. Catling, Ann.Rev.AA **41**, 429 (2003).
10. W. von Bloh, C. Bounama, M. Cuntz, S. Franck, A&A **476**, 1365 (2007).
11. F. Selsis, J. F. Kasting, B. Levrard, J. Paillet, I. Ribas, X. Delfosse, A&A **476,** 1373 (2007).
12. H. R. A. Jones, J. Rayner, L. Ramsey, D. Henry, B. Dent, D. Montgomery, A. Vick, D. Ives, I. Egan, D. Lunney, P. Rees, A. Webster, C. Tinney, M. Liu, Proc SPIE **7014,** 70140Y1, (2008).
13. A. Reiners, J.L. Bean, K.F. Huber, S. Dreizler, A. Seifahrt, S. Czesla, arXiv:0909.0002v1 (2009).
14. J. D. Kirkpatrick, D. M. Kelly, G. H. Rieke, J. Liebert, F. Allard, R. Wehrse, ApJ **402,** 643 (1993).
15. J. Ge, D. McDavitt, B. Zhao, S. Mahadevan, C. DeWitt, S. Seager, Proc. SPIE **6269**, 62691D 1, (2006).
16. RECONS Census of Objects Nearer than 10 Parsecs as of 1/1/2009, http://www.recons.org/
17. G. W. Marcy, R. B. Butler, PASP **104**, 270 (1992).
18. G. W. Marcy, P. R. Butler, D. Fischer, S. Vogt, J. T. Wright, C. G. Tinney, . R. A. Jones, Prog. Theo. Phys. Suppl. **158**, 24 (2005).
19. F. D. D'Amato, E. Oliva, L. Origlia, Proc SPIE, **6269**, 62695E1 (2006).
20. P. Martin, R. Holdsworth, Spectroscopy Europe, **16-5,** 8 (2004).
21. S. Mahadevan, and J. Ge., ApJ **692**, 1590 (2009).
22. 42.  N. C. Santos, F. Bouchy, M. Mayor, F. Pepe, D. Queloz, S. Udry, C. Lovis, M. Bazot, W. Benz, J.-L. Bertaux, G. Lo Curto, X. Delfosse, C. Mordasini, D. Naef. J.-P. Sivan, S. Vauclair, A&A **426**, L19 (2004).
23. T. T. Murphy, P. Tzanavaris, J. K. Webb, C. Lovis, MNRAS **378**, 221 (2007).
24. F. Kerber, F. Saitta, P. Bristow, Messenge **29**, 21 (2007).
25. H. U. Käufl, *Globular Clusters – Guides to Galaxies* (Springer, 2008).
26. S. Osterman, S. Diddams, M. Beasley, C. Froning, L. Hollberg, P. MacQueen, V. Mbele, A. Weiner, Proc. SPIE **6693** G1(2007).
27. M. Murphy, Th. Udem, R. Holzwarth, A. Sizmann, L. Pasquini, C. Araujo-Hauck, H. Dekker, S. D'Odorico, M. Hischer, T. W. Hänsch, A. Manescau, A., MNRAS **380(2)**, 839 (2007).
28. J. N. Eckstein, A. I. Ferguson, T. W. Hänsch, Phys. Rev. Lett. **40**, 847 (1978).
29. Th. Udem, R. Holzwarth, T. Hänsch, Nature **416**, 233 (2002).
30. D. J. Jones, S. A. Diddams, K. K. Ranka, A. Stentz, R. S. Windeler, J. L. Hall, S. T. Cundiff, S.T., Science **288**, 635 (2000).
31. L. –S. Ma, Z. Bi, A. Bartels, L. Robertsson, M. Zucco, R. S. Windeler, G. Wilpers, C. Oates, L. Hollberg, S. A. Diddams, Science **303**, 1843 (2004).
32. F. Quinlan, Y. Jiang, D. Braje, S. Osterman, S. Diddams, 2009 IEEE Photonics Society Annual Meeting (Belek-Antalya, Turkey), paper TuG3
33. T. Sizer, II, J Quantum Elect, **25**, 97 (1989).
34. D. A. Braje, M. S. Kirchner, S. Osterman, T. Fortier S. A. Diddams, EPJ-D **48**, 57 (2008).